\documentclass[aps,pra,showpacs,twocolumn, longbibliography]{revtex4-1}

\usepackage{graphicx}
\usepackage{dcolumn}
\usepackage{bm}
\usepackage[colorlinks]{hyperref}
\usepackage{amsmath}

\begin{document}


\title{Spectrally resolved Franson interference}

\author{Rui-Bo Jin$^{1 }$}
\author{Zi-Qi Zeng$^{1 }$}
\author{Dan Xu$^{1 }$}
\author{Chen-Zhi Yuan$^{1}$}
\email{chenzhi.yuan@wit.edu.cn}
\author{Bai-Hong Li$^{2}$}
\email{li-baihong@163.com}
\author{You Wang$^{3}$}
\author{Ryosuke Shimizu$^{4}$}
\author{Masahiro Takeoka$^{5}$}
\author{Mikio Fujiwara$^{6}$}
\author{Masahide Sasaki$^{6}$}
\author{Pei-Xiang Lu$^{1}$}

%
\affiliation{$^{1}$Hubei Key Laboratory of Optical Information and  Pattern Recognition, Wuhan Institute of Technology, Wuhan 430205, China}
\affiliation{$^{2}$ Department of Physics, Shaanxi University of Science and Technology, Xi'an 710021, China}
\affiliation{$^{3}$ Southwest Institute of Technical Physics, Chengdu 610041, China}
\affiliation{$^{4}$ University of Electro-Communications, 1-5-1 Chofugaoka, Chofu, Tokyo 182-8585, Japan}
\affiliation{$^{5}$ Keio University, 3-14-1 Hiyoshi, Kohoku, Yokohama, Kanagawa 223-8522, Japan}
\affiliation{$^{6}$ National Institute of Information and Communications Technology , 4-2-1 Nukui-Kitamachi, Koganei, Tokyo 184-8795, Japan}

\date{\today}
\begin{abstract}
Franson interference can be used to test the nonlocal features of energy-time entanglement and has become a standard in quantum physics.
However, most of the previous Franson interference experiments were demonstrated in the time domain, and the spectral properties of Franson interference have not been fully explored.
Here, we theoretically and experimentally demonstrate spectrally resolved Franson interference using biphotons with different correlations, including positive correlation, negative correlation, and non-correlation.
It is found that the joint spectral intensities of the biphotons can be modulated along both the signal and idler directions, which has potential applications in generating high-dimensional frequency entanglement and time-frequency grid states.
This work may provide a new perspective for understanding the spectral-temporal properties of the Franson interferometer.
\end{abstract}

\maketitle

\section{Introduction}
Franson interference was proposed in 1989 to test the Bell inequality for position or time, specifically to explore the feasibility of local hidden-variable models using a new optical interferometer \cite{Franson1989}.
In a typical configuration for the Franson interferometer, the signal and idler photons, generated simultaneously, are distributed to different terminals while passing through unbalanced Mach-Zehnder interferometers (UMZIs) inserted in their paths. The signal and idler photons can choose either short or long pathways within the UMZIs. In delayed coincidence measurements, it is convenient to consider only events where both signal and idler photons select either the short or long pathways. Since these two cases can be indistinguishable, they can interfere with each other. Interference fringes can be observed in the coincidence measurements when the optical path-length difference in the UMZI is shorter than the two-photon coherence length of the signal and idler photons. The interference from single photons can be eliminated by setting the optical path-length difference longer than the coherence length of either the signal or idler photons.
Several experiments have modified the Franson interferometer from its original configuration for different purposes \cite{ Cabello2009PRL, Kwiat1990PRA, Mittal2021NP}. For instance, hug-type configurations have been invented to remove the post-selection loophole in the original configuration\cite{Cabello2009PRL}, and single Michelson configurations have been used to make the setup more compact\cite{Kwiat1990PRA,Mittal2021NP}.
 %

%
\begin{figure*}[!thp]
\centering
\includegraphics[width= 0.85\textwidth]{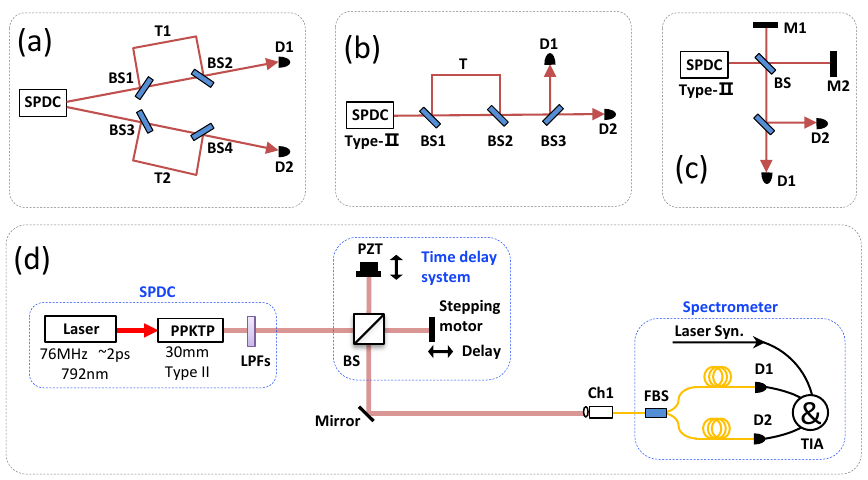}
\caption{ (a) The model of the traditional unfolded Franson interference.
(b) The Mach–Zehnder-type folded Franson interference.
(c) The Michelson-type folded Franson interference.
(d) The experimental setup based on (c). LPFs = long-pass filters, PZT = piezoelectric motor, BS = beam splitter, FBS = fiber beam splitter, D = detector, TIA = time interval analyzer.
}
\label{setup}
\end{figure*}

Numerous experiments have been conducted to observe Franson interference, which has become a standard tool in quantum optics for verifying energy-time or time-bin entanglement \cite{Ou1990PRL,Brendel1991PRL}. In these experiments, various mechanisms have been employed to generate photon pairs, including spontaneous parametric down-conversion (SPDC) processes in bulk crystals with $\chi^{(2)}$ nonlinearity \cite{Ou1990PRL,Brendel1991PRL}, SPDC or spontaneous four-wave mixing (SFWM) processes in waveguides or microresonators with $\chi^{(2)}$ or $\chi^{(3)}$ nonlinearities \cite{Sanaka2001PRL,Ma2020PRL,Grassani2015Optica}, SFWM in atomic ensembles \cite{Park2018PRL}, and cascaded emission in quantum dots (QDs) \cite{Jayakumar2014NC}. The applications of Franson interference range from testing fundamental physical principles \cite{Tittel1998PRL,Stefanov2002PRL} to quantum cryptography \cite{AliKhan2007PRL}, entanglement-based quantum networks \cite{Sun2017Optica}, and quantum imaging \cite{Gao2019APL}.
However, most of the previous Franson interference experiments have focused on time-resolved measurements, and it is expected that a spectrally-resolved configuration would provide new capabilities.

Spectrally-resolved interferometers create interference fringes with different frequency components separated spatially or temporally. These interferometers have already been employed in measuring the linear and nonlinear dielectric properties of materials \cite{Tokunaga1992OL}, coherently controlling ultrafast carrier dynamics in semiconductor nanostructures \cite{Heberle1995PRL}, measuring laser-generated shock waves in metal thin films \cite{Gahagan2000PRL}, and studying the dynamics of ultrashort laser-produced plasma \cite{Salieres1999PRL}. In the field of quantum optics, frequency-resolved Hong-Ou-Mandel (HOM) interference has been demonstrated \cite{Jin2015OE, Jin2016QST, Orre2019PRL,YepizGraciano2020PR,Merkouche2022PRL} and used in entanglement swapping of energy-time entanglement \cite{Merkouche2022PRL} and quantum optical coherence tomography \cite{YepizGraciano2020PR}.

In this article, we theoretically and experimentally demonstrate a spectrally resolved Franson interferometer. In theory, we confirm that a folded Franson interferometer can achieve the same performance as the original Franson interference. We compare time-resolved and spectrally resolved interferograms for biphotons with positive correlation, negative correlation, and non-correlation. In the experiment, we measure the spectrally resolved interferograms of biphotons generated by SPDC under different time delays.
We find that the joint spectral intensities of the biphoton can be modulated along both the signal and idler directions. Additionally, we observe that the spectrally resolved interferograms remain clear even when the time-resolved interferogram disappears.

\section{Theory and simulation}
%
\begin{figure*}[!thp]
\centering
\includegraphics[width= 0.98\textwidth]{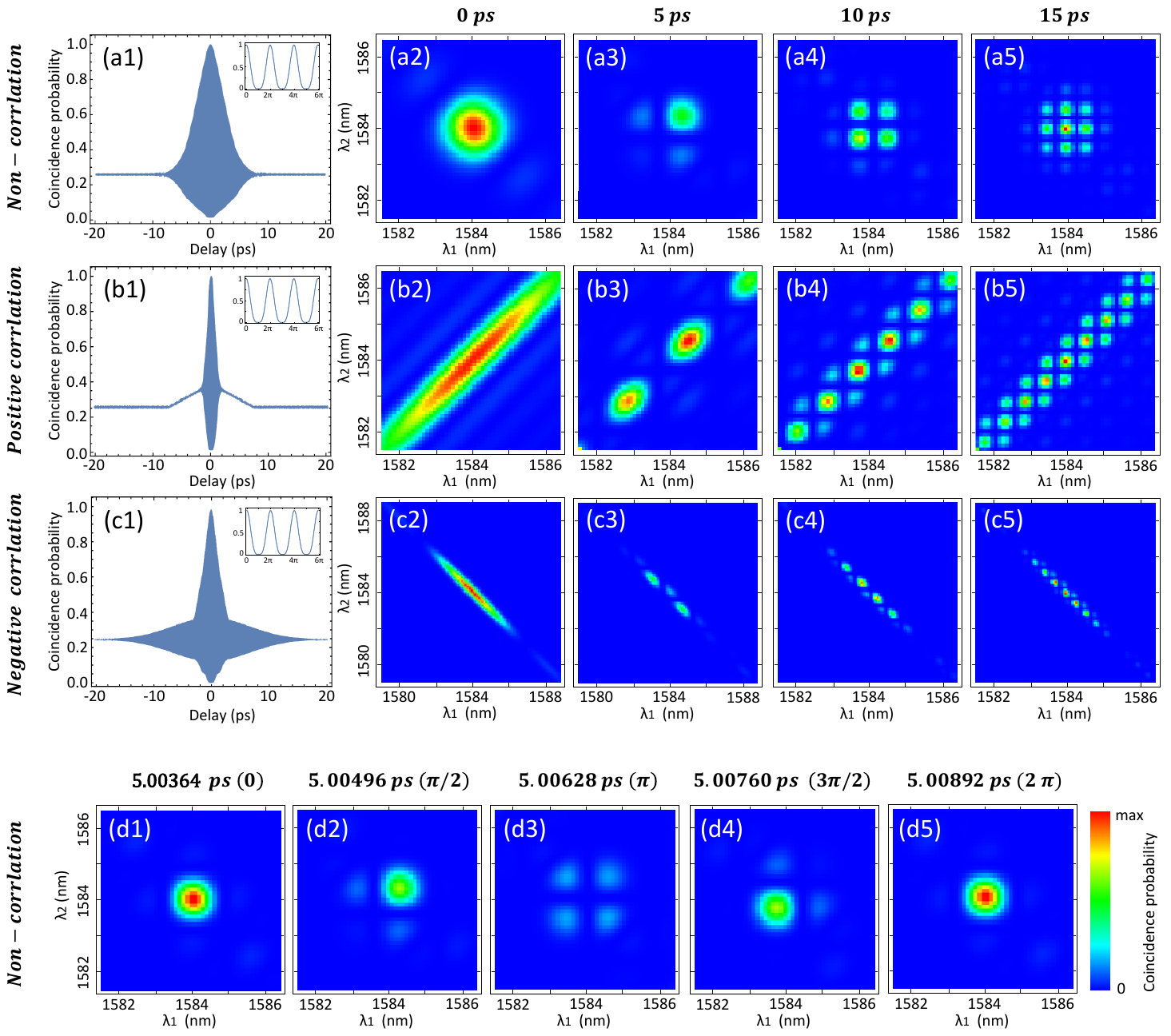}
\caption{ The first, second, and third rows display the simulation results of spectrally non-correlated, positively correlated, and negatively correlated biphotons, respectively.
(a1, b1, c1) are the simulated coincidence probability $P(\tau)$ as a function of the delay $\tau$. The insets in (a1, b1, c1) show  $P(\tau)$ within a time duration of 0 to 15.84 fs, corresponding to a phase delay of 0-6$\pi$.
The spectral distributions $S(\omega_s, \omega_i, \tau)$ at different time delays are represented by (a2-a5), (b2-b5), and (c2-c5) for 0 ps, 5 ps, 10 ps, and 15 ps, respectively. On the other hand, (d1-d5) illustrate $S(\omega_s, \omega_i, \tau)$ for spectrally non-correlated biphotons with a time delay ranging from 5.00364 ps to 5.00892 ps, corresponding to phases from 0 to $2\pi$.
}
\label{simulation}
\end{figure*}

The two-photon state from an SPDC process can be described as:
\begin{equation}\label{eq1}
\left| \psi \right\rangle = \int_0^{\infty}{\int_0^\infty {d\omega _s d\omega _i } } f(\omega _s ,\omega _i )\hat a_s^\dag (\omega _s )\hat a_i^\dag (\omega _i )\left| {00} \right\rangle,
\end{equation}
where $\omega$ is the angular frequency, $\hat a^\dag$ is the creation operator, and the subscripts $s$ and $i$ denote the signal and idler photons from SPDC, respectively. $f(\omega _s ,\omega _i )$ represents the joint spectral amplitude of the signal and idler photons.

As  calculated in the Appendix, the coincidence probability $P_{0}(\tau )$ in the traditional unfolded Franson interference (with the setup in   Fig.\,\ref{setup}(a)) is given by:
\begin{equation}\label{eq2}
\begin{array}{lll}
P_0(\tau)&=&\int_0^{\infty} \int_0^{\infty} d \omega_s d \omega_i\left|f\left(\omega_s, \omega_i\right)\right|^2 \\
 &&\times \left[1+\cos\left(\omega_s \tau \right)\right]\left[1+\cos\left(\omega_i \tau \right)\right],  \\
 \end{array}
\end{equation}
where $\tau$ is the optical path delay between the long  and the short arms.
For a folded Franson interference with the setup in   Fig.\,\ref{setup}(b) or (c),  the coincidence probability $P(\tau)$ is given by:
\begin{equation}\label{eq3}
\begin{array}{lll}
P(\tau)  &=&\frac{1}{8} \int_0^{\infty} \int_0^{\infty} d \omega_s d \omega_i \left|f\left(\omega_s, \omega_i\right)\right|^2 \\
&& \times\left[1+\cos\left(\omega_s \tau\right)\right]\left[1+\cos\left(\omega_i \tau\right)\right]. \\
 \end{array}
\end{equation}
The joint spectral correlation  $S(\omega_s, \omega_i, \tau)$ at different delay positions can be calculated as:
\begin{equation}\label{eq4}
\begin{aligned}
S(\omega_s, \omega_i, \tau)  = \left|f\left(\omega_s, \omega_i\right)\right|^2 
\left[1+\cos\left(\omega_s \tau\right)\right]\left[1+\cos\left(\omega_i \tau\right)\right].\\
\end{aligned}
\end{equation}
Eq.\,(\ref{eq2}) and Eq.\,(\ref{eq3}) have a similar form, indicating that the folded Franson interference can achieve the same performance as the original Franson interference.

By using  Eq.\,(\ref{eq3}) and Eq.\,(\ref{eq4}) , we can simulate $P(\tau)$ and $S(\omega_s, \omega_i, \tau)$ at different delays $\tau$ and with different spectral distributions  $f\left(\omega_s, \omega_i\right)$, as shown in Fig.\,\ref{simulation}.
The first row presents the case of spectrally non-correlated biphotons. This calculation is performed using a 30-mm-long PPKTP crystal and a pump laser with a Gaussian distribution. The laser has a center wavelength of 792 nm and a full width at half maximum (FWHM) of 0.40 nm.
The second row is the case of positively correlated biphotons, which are calculated using a 50-mm-long PPKTP crystal and a pump laser with an FWHM of 2.35 nm.
The third row shows the case of negatively correlated biphotons, which are calculated using a 10-mm-long PPKTP crystal and a pump laser with an FWHM of 0.12 nm.

Fig.\,\ref{simulation}(a1, b1, c1) displays  the coincidence probability $P(\tau)$ as a function of delay, while the corresponding $|f (\omega_1, \omega_2 )|^2$ is shown in Fig.\,\ref{simulation}(a2, b2, c2), respectively. 
Within the range of -20 ps to 20 ps, the envelope of the coincidence probability exhibits distinct variations for biphotons with different correlations. However, between 0 and 15.84 fs, the interference patterns remain consistent, as illustrated by the insets in Fig.\,\ref{simulation}(a1, b1, c1).

The spectral distribution $S (\omega_s, \omega_i, \tau  )$ with different correlations at 0 ps, 5 ps, 10 ps, and 15 ps are depicted in  Fig.\,\ref{simulation}(a2-a5), (b2-b5), and (c2-b5).
Notably, an increase in delay leads to a greater separation of spectral modes into multiple components. This phenomenon can be effectively explained by Eq.\,(\ref{eq3}).
To facilitate a comparison of the spectral distribution at different phases, Fig.\,\ref{simulation}(d1-d5) illustrates $S (\omega_s, \omega_i, \tau )$ at phase differences of 0, $\pi/2$, $\pi$, $3\pi/2$, and $2\pi$.
We can observe  that the mode number changes gradually from 1 mode to 4 modes,  and then returns to  1 mode.

\section{Experiment and results}
%
%
\begin{figure*}[!thp]
\centering
\includegraphics[width= 0.98\textwidth]{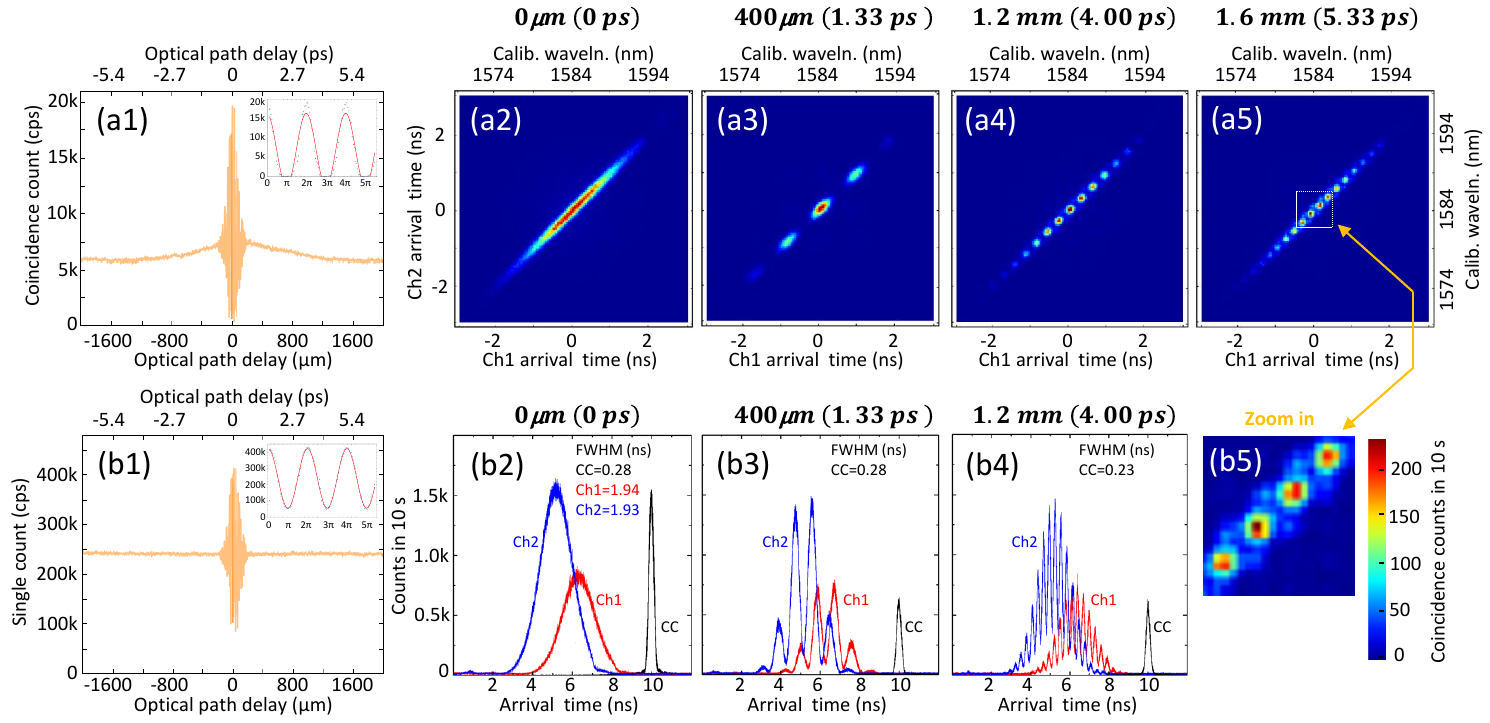}
\caption{ Experimental results. (a1, b1) The measured coincidence (single) counts   as a function of the time delay scanned with a stepping motor,  with a  step of 4 $\mu$m. The insets show the  measured coincidence (single) counts by scanning a PZT with a  step of 40 nm. 
(a2-a5) The measured JSIs at the delay position of 0 ps, 1.33 ps, 4.00 ps, and 5.33 ps, respectively. The accumulation time is 10 seconds for each figure. 
(b2-b4) The time-of-arrival measurement for single count of channel 1 (SC1,  in red), single count of channel 2 (SC2, in blue), and coincidence counts (CC, in black) at  0 ps, 1.33 ps, and 4.00 ps.
(b5) is an enlarged view of the center section of (a5).
}
\label{fig3}
\end{figure*}
%
The experimental setup is shown in  Fig.\,\ref{setup}(d). 
Laser pulses with a temporal width of around 2 ps and a center wavelength of 792 nm were utilized to pump a 30-mm-long periodically poled KTiOPO$_4$ (PPKTP) crystal.
The PPKTP crystal was type-II phase matched (y$\to$y+z), and the signal and idler photons generated from the SPDC process were orthogonally polarized \cite{Jin2022APL}.
After filtering by the long-path filters, the biphotons were sent to a time delay system, which consisted of a beamsplitter (BS), a PZT, and a stepping motor.
Then, the photons were coupled into a  fiber beamsplitter, which was connected to a fiber spectrometer .
The fiber spectrometer consisted of  two 7.5-km-long SMFs, two SNSPDs, one synchronization signal from the laser,  and a TIA \cite{Jin2016QST}.
The dispersion of the SMFs was calibrated as 27.3 ps/km/nm at 1584 nm. 
Considering an estimated 100 ps FWHM jitter of the detection system, the resolution of this fiber spectrometer was calculated to be 0.5 nm.

The measured coincidence counts as a function of optical path delay are shown in Fig.\,\ref{fig3}(a1).
The main figure was obtained by scanning the stepping motor with a step length of 4 $\mu$m. The FWHM of the upper envelope is 0.56 ps
The insert in Fig.\,\ref{fig3}(a1) was obtained by scanning a PZT with a step length of 40 nm. The visibility is 99.90\% $\pm$ 0.00\%, indicating a high indistinguishability of the signal and idler photons.
The main figure and insert in Fig.\,\ref{fig3}(a1) are consistent with the simulation results in  Fig.\,\ref{simulation}(b1).

Fig.\,\ref{fig3}(a2-a5) shows the measured JSI at 0 ps, 1.33 ps, 4 ps, and 5.33 ps respectively.
It can be observed that with the increase of time delay, the mode number increases.
The mode numbers in Fig.\,\ref{fig3}(a2-a5) are 1, 3, 6, and 15,  respectively.
Fig.\,\ref{fig3}(b5) is  an enlarged view of (a5), and it is clear that the modes are separated in both the horizontal and vertical directions..

We also measured the single counts at the same time as the coincidence measurement, as shown in Fig.\,\ref{fig3}(b1).
The single counts have a constant baseline, which is different from the varying baseline in the coincidence counts in Fig.\,\ref{fig3}(a1).
The insert in Fig.\,\ref{fig3}(b1) shows the single counts,  obtained by scanning the PZT,  and the visibility is 
79.16\% $\pm$ 0.05\%.
We also measured the time-of-arrival (TOA)  of channel 1 (ch1) and channel 2 (ch2) in Fig.\,\ref{fig3}(b2-b4).
It can be observed that with the increase of time delay, the single peak in the single counts evolves into multiple peaks. However, the peak in the TOA of the coincidence counts remains a single peak.
This is caused by the fact that the TOA  of single counts is obtained by projecting the JSI data onto the horizontal and vertical axes, while the TOA of the coincidence counts is obtained by projecting the  JSI data onto the anti-diagonal line, i.e., the line of $\omega_s - \omega_i$.

\section{Discussion}
There are two types of coherence time for biphotons: the sum-frequency coherence time and the difference-frequency coherence time \cite{Jin2018PRAppl, MacLean2018PRA}.
The sum-frequency coherence time is determined by the pump laser and can be tested in the NOON state interference.
The difference-frequency coherence time is determined by the phase-matching condition of the nonlinear crystal and can be tested in the HOM interference \cite{Jin2018Optica}.
For single photons, the coherence time is determined by the projection of joint temporal distributions onto the signal or idler direction.
In traditional Franson interference, the sum-frequency coherence time is much longer than the coherence time of single photons.

The spectrally resolved measurement has been previously investigated in HOM interference \cite{Gerrits2015,Jin2015OE, Chen2023PRAppl}, modified HOM interference\cite{Li2023PRA}, NOON state interference  \cite{Jin2021arXiv}, 
and also demonstrated in the characterization of time-energy entangled state \cite{MacLean2018PRA, MacLean2019PRA}.
It can be observed that even when there are no interference patterns in the time domain, the interference patterns in the spectral domain are still very clear.
The spectral measurement in interference can be fundamentally understood as a tool for temporal filtering, which increases the coherence time of the photons by filtering. 
The JSI measured in quantum interference is helpful and is complementary to the measurement of temporal interference.

Since the joint spectral intensities of the biphotons can be modulated along both the signal and idler directions, it is possible to generate high-dimensional entangled states (entangled qudits) and time-frequency grid states using spectrally resolved Franson interference. 
As demonstrated in Fig.\,\ref{fig3}(a4), this is indeed a kind of entangled qudits \cite{Yang2023OL}.
For example, the state generated in Fig.\,\ref{simulation}(a5) is a time-frequency grid state \cite{Fabre2020PRA}, 
which can be used to implement measurement-based quantum error correction in fault-tolerant quantum computing using time-frequency continuous variables \cite{Gottesman2001PRA,Menicucci2014PRL,Baragiola2019PRL}.

\section{Conclusion}
In summary, we have theoretically and experimentally demonstrated spectrally resolved Franson interference using biphotons with different correlations. 
The joint spectral intensities of the biphotons were measured at different delay positions in an Franson interference.
It can be observed that even when there are no interference patterns in the time domain, the interference patterns in the spectral domain are still very clear.
This work provides a new perspective by considering the joint spectral distribution to understand the spectral-temporal properties in Franson interference. Furthermore, this approach can be used to generate high-dimensional entangled states and time-frequency grid states.

\section*{Acknowledgments}
This work was supported by the National Natural Science Foundations of China (Grant Numbers 92365106, 12074309,  and 12074299), and the Natural Science Foundation of Hubei Province (2022CFA039).

\onecolumngrid
\clearpage
\renewcommand\thefigure{A\arabic{figure}}
\setcounter{figure}{0}

\setcounter{equation}{0}
\renewcommand\theequation{A\arabic{equation}}

\newpage

\subsection*{Appendix 1: Calculation of standard Franson interference}

%
\begin{figure*}[!thp]
\centering
\includegraphics[width= 0.55\textwidth]{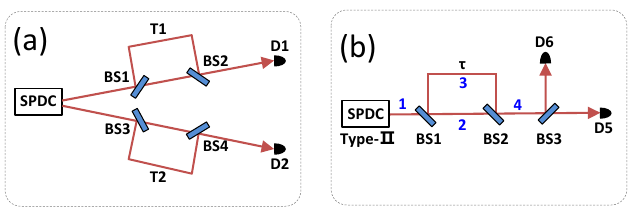}
\caption{ (a) The setup of an unfolded Franson interferometer. (b) The setup of a folded Franson interferometer.
}
\label{SupplementalFig}
\end{figure*}

In this section, we deduce the equations for the Franson interference using multi-mode theory. The setup of the Franson interference is shown in Fig.\,\ref{SupplementalFig} (a).
The two-photon state from a spontaneous parametric down-conversion (SPDC) process can be described as
\begin{equation}\label{eqA1}
\left| \psi  \right\rangle  = \int_0^{\infty}{\int_0^\infty  {d\omega _s d\omega _i } } f(\omega _s ,\omega _i )\hat a_s^\dag  (\omega _s )\hat a_i^\dag  (\omega _i )\left| {00} \right\rangle,
\end{equation}
where $\omega$ is the angular frequency; $\hat a^\dag$ is the creation operator and the subscripts $s$ and $i$ denote the signal and idler photons from SPDC, respectively; $f(\omega _s ,\omega _i )$ is the joint spectral amplitude of the signal and idler photons.
%
%

The detection field operators of detector 1 (D1) and detector 2 (D2) are
\begin{equation}\label{eqA2}
\hat E_1^{( + )} (t_1 ) = \frac{1}{{2 }}\int_0^\infty  {d\omega _1 } \hat a_1 (\omega _1 )e^{ - i\omega _1 t_1 },
\end{equation}
\begin{equation}\label{eqA3}
\hat E_2^{( + )} (t_2 ) = \frac{1}{{2} }\int_0^\infty  {d\omega _2 \hat a_2 (\omega _2 )} e^{ - i\omega _2 t_2 },
\end{equation}
where the subscripts $1$ and $2$ denote the photons detected by D1 and D2, respectively.
The transformation rule after the delay times $T_1$ and $T_2$ is
\begin{equation}\label{eqA4}
\hat{a}_1\left(\omega_1\right)=\frac{1}{\sqrt{2}}\left[\hat{a}_s\left(\omega_1\right)+\hat{a}_s\left(\omega_1\right) e^{-i \omega_1 T_1}\right],
\end{equation}
\begin{equation}\label{eqA5}
\hat{a}_2\left(\omega_2\right)=\frac{1}{\sqrt{2}}\left[\hat{a}_i\left(\omega_2\right)+\hat{a}_i\left(\omega_2\right) e^{-i \omega_2 T_2}\right].
\end{equation}

So, we can rewrite the field operators as
\begin{equation}\label{eqA6}
\begin{array}{lll}
\hat{E}_1^{(+)}\left(t_1\right)=\frac{1}{2 \sqrt{2 \pi}} \int_0^{\infty} d \omega_1\left[\hat{a}_s\left(\omega_1\right)+\hat{a}_s\left(\omega_1\right) e^{-i \omega_1 T_1}\right] e^{-i \omega_1 t_1},\\
 \end{array}
\end{equation}
and
\begin{equation}\label{eqA7}
\begin{array}{lll}
\hat{E}_2^{(+)}\left(t_2\right)=\frac{1}{2 \sqrt{2 \pi}} \int_0^{\infty} d \omega_2\left[\hat{a}_i\left(\omega_2\right)+\hat{a}_i\left(\omega_2\right) e^{-i \omega_2 T_2}\right] e^{-i \omega_2 t_2}. \\
 \end{array}
 \end{equation}

The coincidence probability $P(\tau )$, which is also the second-order correlation function $G_2(\tau)$,  can be expressed as
\begin{equation}\label{eq0}
P(\tau ) \equiv G_2(\tau)    = \int_{-\infty}^{\infty}  {\int_{-\infty}^{\infty} {dt_1 dt_2 } } \left\langle {\psi \left| {\hat E_1^{( - )} \hat E_2^{( - )} \hat E_2^{( + )} \hat E_1^{( + )} } \right|\psi } \right\rangle.
\end{equation}

First of all, consider $\hat E_2^{( + )} \hat E_1^{( + )}$,
\begin{equation}\label{eq0}
\begin{array}{l}
 \begin{aligned}
\hat{E}_2^{(+)} \hat{E}_1^{(+)} & =\frac{1}{8 \pi} \int_0^{\infty} \int_0^{\infty} d \omega_1 d \omega_2\left[\hat{a}_s\left(\omega_1\right)+\hat{a}_s\left(\omega_1\right) e^{-i \omega_1 T_1}\right]\left[\hat{a}_i\left(\omega_2\right)+\hat{a}_i\left(\omega_2\right) e^{-i \omega_2 T_2}\right] e^{-i \omega_1 t_1} e^{-i \omega_2 t_2} \\
& =\frac{1}{8 \pi} \int_0^{\infty} \int_0^{\infty} d \omega_1 d \omega_2 \hat{a}_s\left(\omega_1\right) \hat{a}_i\left(\omega_2\right)\left[1+e^{-i \omega_1 T_1}+e^{-i \omega_2 T_2}+e^{-i\left(\omega_1 T_1+\omega_2 T_2\right)}\right] e^{-i \omega_1 t_1} e^{-i \omega_2 t_2} \\
& =\frac{1}{8 \pi} \int_0^{\infty} \int_0^{\infty} d \omega_1 d \omega_2 \hat{a}_s\left(\omega_1\right) \hat{a}_i\left(\omega_2\right)\left(1+e^{-i \omega_1 T_1}\right)\left(1+e^{-i \omega_2 T_2}\right) e^{-i \omega_1 t_1} e^{-i \omega_2 t_2}.
\end{aligned}\\
 \end{array}
\end{equation}

Then, consider $\hat{E}_2^{(+)} \hat{E}_1^{(+)}|\psi\rangle $
\begin{equation}\label{eq0}
\begin{array}{l}
 \begin{aligned}
\hat{E}_2^{(+)} \hat{E}_1^{(+)}|\psi\rangle & =\frac{1}{8 \pi} \int_0^{\infty} \int_0^{\infty} d \omega_1 d \omega_2 \hat{a}_s\left(\omega_1\right) \hat{a}_i\left(\omega_2\right)\left(1+e^{-i \omega_1 T_1}\right)\left(1+e^{-i \omega_2 T_2}\right) e^{-i \omega_1 t_1} e^{-i \omega_2 t_2} \\
& \times \int_0^{\infty} d \omega_s \int_0^{\infty} d \omega_i f\left(\omega_s, \omega_i\right) \hat{a}_s^{\dagger}\left(\omega_s\right) \hat{a}_i^{\dagger}\left(\omega_i\right)|00\rangle \\
& =\frac{1}{8 \pi} \int_0^{\infty} \int_0^{\infty} \int_0^{\infty} \int_0^{\infty} d \omega_1 d \omega_2 d \omega_s d \omega_i f\left(\omega_s, \omega_i\right) \delta\left(\omega_1-\omega_s\right) \delta\left(\omega_2-\omega_i\right) \\
& \times\left(1+e^{-i \omega_1 T_1}\right)\left(1+e^{-i \omega_2 T_2}\right) e^{-i \omega_1 t_1} e^{-i \omega_2 t_2}|00\rangle \\
& =\frac{1}{8 \pi} \int_0^{\infty} \int_0^{\infty} d \omega_1 d \omega_2 f\left(\omega_1, \omega_2\right)\left(1+e^{-i \omega_1 T_1}\right)\left(1+e^{-i \omega_2 T_2}\right) e^{-i \omega_1 t_1} e^{-i \omega_2 t_2}|00\rangle.
\end{aligned}\\
 \end{array}
\end{equation}

In the above calculation, the equations of $\hat a_s (\omega _1 )\hat a_s^\dag  (\omega _s )  \left| {0} \right\rangle = \delta (\omega _1  - \omega _s ) \left| {0} \right\rangle$  and $\hat a_i (\omega _2 )\hat a_i^\dag  (\omega _i )\left| {0} \right\rangle = \delta (\omega _2  - \omega _i ) \left| {0} \right\rangle$ are used.

Then,
\begin{equation}\label{eq0}
\begin{array}{lll}
 \begin{aligned}
\left\langle\psi\left|\hat{E}_1^{(-)} \hat{E}_2^{(-)} \hat{E}_2^{(+)} \hat{E}_1^{(+)}\right| \psi\right\rangle & =\frac{1}{64 \pi^2} \int_0^{\infty} \int_0^{\infty} d \omega_1^{\prime} d \omega_2^{\prime} f^*\left(\omega_1^{\prime}, \omega_2^{\prime}\right)\left(1+e^{i \omega_1^{\prime} T_1}\right)\left(1+e^{i \omega_2^{\prime} T_2}\right) e^{i \omega_1^{\prime} t_1} e^{i \omega_2^{\prime} t_2} \\
& \times \int_0^{\infty} \int_0^{\infty} d \omega_1 d \omega_2 f\left(\omega_1, \omega_2\right)\left(1+e^{-i \omega_1 T_1}\right)\left(1+e^{-i \omega_2 T_2}\right) e^{-i \omega_1 t_1} e^{-i \omega_2 t_2}.
\end{aligned}  \\
 \end{array}
\end{equation}

Finally,
\begin{equation}\label{eq0}
\begin{array}{lll}
\begin{aligned}
P(\tau) & =\int_{-\infty}^{\infty} \int_{-\infty}^{\infty} d t_1 d t_2\left\langle\psi\left|\hat{E}_1^{(-)} \hat{E}_2^{(-)} \hat{E}_2^{(+)} \hat{E}_1^{(+)}\right| \psi\right\rangle \\
& =\frac{1}{64 \pi^2} \int_0^{\infty} \int_0^{\infty} d t_1 d t_2 \int_0^{\infty} \int_0^{\infty} d \omega_1^{\prime} d \omega_2^{\prime} f^*\left(\omega_1^{\prime}, \omega_2^{\prime}\right)\left(1+e^{i \omega_1^{\prime} T_1}\right)\left(1+e^{i \omega_2^{\prime} T_2}\right) e^{i \omega_1^{\prime} t_1} e^{i \omega_2^{\prime} t_2} \\
& \times \int_0^{\infty} \int_0^{\infty} d \omega_1 d \omega_2 f\left(\omega_1, \omega_2\right)\left(1+e^{-i \omega_1 T_1}\right)\left(1+e^{-i \omega_2 T_2}\right) e^{-i \omega_1 t_1} e^{-i \omega_2 t_2}.
\end{aligned} \\
 \end{array}
\end{equation}

By utilizing $\frac{1}{2 \pi} \int_{-\infty}^{\infty} e^{-i(\omega-\omega^{\prime}) t} d t=\delta(\omega-\omega^{\prime})$, the above equation can be further simplified as \begin{equation}\label{eq0}
\begin{array}{lll}
\begin{aligned}
P(\tau) & =\int_{-\infty}^{\infty}\int_{-\infty}^{\infty} d t_1 d t_2\left\langle\psi\left|\hat{E}_1^{(-)} \hat{E}_2^{(-)} \hat{E}_2^{(+)} \hat{E}_1^{(+)}\right| \psi\right\rangle \\
& =\frac{1}{16} \int_0^{\infty} \int_0^{\infty} \int_0^{\infty} \int_0^{\infty} d \omega_1 d \omega_2 d \omega_1^{\prime} d \omega_2^{\prime} \delta\left(\omega_1-\omega_1^{\prime}\right) \delta\left(\omega_2-\omega_2^{\prime}\right) f\left(\omega_1, \omega_2\right) \\
& \times\left(1+e^{-i \omega_1 T_1}\right)\left(1+e^{-i \omega_2 T_2}\right) f^*\left(\omega_1^{\prime}, \omega_2^{\prime}\right)\left(1+e^{i \omega_1^{\prime} T_1}\right)\left(1+e^{i \omega_2^{\prime} T_2}\right) e^{i \omega_1^{\prime} t_1} e^{i \omega_2^{\prime} t_2} \\
& =\frac{1}{16} \int_0^{\infty} \int_0^{\infty} d \omega_1 d \omega_2 f\left(\omega_1, \omega_2\right) f^*\left(\omega_1, \omega_2\right)\left(1+e^{-i \omega_1 T_1}\right)\left(1+e^{-i \omega_2 T_2}\right)\left(1+e^{i \omega_1 T_1}\right)\left(1+e^{i \omega_2 T_2}\right)  \\
& =\frac{1}{4} \int_0^{\infty} \int_0^{\infty} d \omega_1 d \omega_2\left|f\left(\omega_1, \omega_2\right)\right|^2\left[1+\operatorname{cos}\left(\omega_1 T_1\right)\right]\left[1+\operatorname{cos}\left(\omega_2 T_2\right)\right]. 
\end{aligned}  \\
 \end{array}
\end{equation}

If the delay of the two paths is now equal, then:
\begin{equation}\label{eq0}
\begin{array}{lll}
\begin{aligned}
P(\tau)=\frac{1}{4} \int_0^{\infty} \int_0^{\infty} d \omega_1 d \omega_2\left|f\left(\omega_1, \omega_2\right)\right|^2\left[1+\operatorname{cos}\left(\omega_1 T\right)\right]\left[1+\operatorname{cos}\left(\omega_2 T\right)\right]  \\
\end{aligned}
 \end{array}.
\end{equation}

Next, calculate the count probability of the single count and take detector 1 as an example. Assuming that the biphoton state produced in the SPDC process is separable, then the single photon state passing through the path of T1 is:

\begin{equation}\label{eq0}
\begin{array}{lll}
\begin{aligned}
\left| \psi  \right\rangle_s  = \int_0^{\infty}    {d{\omega _s}} f({\omega _s})\hat a_s^\dag ({\omega _s})\left| 0 \right\rangle. 
\end{aligned}
 \end{array}
\end{equation}

Similarly, the detector operator is:
\begin{equation}\label{eq0}
\begin{array}{lll}
\begin{aligned}
\hat E_1^{( + )}({t_1}) = \frac{1}{{\sqrt {2\pi } }}\int_0^{\infty}  {d{\omega _1}} {\hat a_1}({\omega _1}){e^{ - i{\omega _1}{t_1}}}.
\end{aligned}
 \end{array}
\end{equation}
The transformation rule after the delay time $T_1$ is
\begin{equation}\label{eq0}
\begin{array}{lll}
\begin{aligned}
{\rm{  }}{\hat a_1}({\omega _1}) = \frac{1}{2}\left[ {{{\hat a}_s}({\omega _1}) + {{\hat a}_s}({\omega _1}){e^{ - i{\omega _1}{T_1}}}} \right].
\end{aligned}
 \end{array}
\end{equation}

So, we can rewrite the field operators as
\begin{equation}\label{eqA6}
\begin{array}{lll}
\begin{aligned}
\hat E_1^{( + )}({t_1}) = \frac{1}{{2\sqrt {2\pi } }}\int_0^{\infty}  {d{\omega _1}} [{\hat a_s}({\omega _1}) + {\hat a_s}({\omega _1}){e^{ - i{\omega _1}{T_1}}}]{e^{ - i{\omega _1}{t_1}}}\\
\end{aligned}.
 \end{array}
\end{equation}

The single count probability $P_{SC}(\tau )$, can be expressed as
\begin{equation}\label{eq0}
P_{SC}(\tau ) = {\int_{-\infty}^{\infty} {d{t_1}} } \left\langle {\psi \left| {\hat E_1^{( - )}\hat E_1^{( + )}} \right|\psi } \right\rangle.
\end{equation}

Firstly, considering $\hat E_1^{( + )}\left| \psi  \right\rangle$
\begin{equation}\label{eq0}
\begin{array}{l}
 \begin{aligned}
\hat E_1^{( + )}\left| \psi  \right\rangle & = \frac{1}{{2\sqrt {2\pi } }}\int_0^{\infty}  {d{\omega _1}{{\hat a}_s}({\omega _1})} [1 + {e^{ - i{\omega _1}{T_1}}}]{e^{ - i{\omega _1}{t_1}}} \times \int_0^{\infty}  {d{\omega _s}} f({\omega _s})\hat a_s^\dag ({\omega _s})\left| 0 \right\rangle \\
& = \frac{1}{{2\sqrt {2\pi } }}\int_0^{\infty}  {d{\omega _1}} f({\omega _1})[1 + {e^{ - i{\omega _1}{T_1}}}]{e^{ - i{\omega _1}{t_1}}}\left| 0 \right\rangle. 
 \end{aligned}
\end{array}
\end{equation}
In the above calculation, the equations of $\hat a_s (\omega _1 )\hat a_s^\dag  (\omega _s )  \left| {0} \right\rangle = \delta (\omega _1  - \omega _s ) \left| {0} \right\rangle$ are used.

Then,
\begin{equation}\label{eqA6}
\begin{array}{lll}
\left\langle {\psi \left| {\hat E_1^{( - )}\hat E_1^{( + )}} \right|\psi } \right\rangle  = \frac{1}{{8\pi }}\int_0^{\infty}  {d\omega _1^,} \mathop f\nolimits^* (\omega _1^,)[1 + {e^{i\omega _1^,{T_1}}}]{e^{i\omega _1^,{t_1}}} \times \int_0^{\infty}  {d{\omega _1}} f({\omega _1})[1 + {e^{ - i{\omega _1}{T_1}}}]{e^{ - i{\omega _1}{t_1}}}.
\end{array}
\end{equation}

Finally,
\begin{equation}\label{eqA6}
\begin{array}{lll}
\begin{aligned}
P_{SC}(\tau ) & = \int {d{t_1}} \left\langle {\psi \left| {\hat E_1^{( - )}\hat E_1^{( + )}} \right|\psi } \right\rangle \\
& = \frac{1}{{8\pi }}\int_{-\infty}^{\infty}  {d{t_1}} \int_0^{\infty}  {d\omega _1^,} \mathop f\nolimits^* (\omega _1^,)[1 + {e^{i\omega _1^,{T_1}}}]{e^{i\omega _1^,{t_1}}} \times \int_0^{\infty}  {d{\omega _1}} f({\omega _1})[1 + {e^{ - i{\omega _1}{T_1}}}]{e^{ - i{\omega _1}{t_1}}}\\
& = \frac{1}{4}\int_0^{\infty}  {\int_0^{\infty}  {d{\omega _1}d\omega _1^,} }  f({\omega _1})\mathop f\nolimits^* (\omega _1^,)[1 + {e^{ - i{\omega _1}{T_1}}}][1 + {e^{i\omega _1^,{T_1}}}] \delta(\omega-\omega^{\prime})\\
& = \frac{1}{4}\int_0^{\infty}  { {d{\omega _1}} } {\left| {f({\omega _1})[1 + {e^{ - i{\omega _1}{T_1}}}]} \right|^2}\\
& = \frac{1}{2}\int_0^{\infty}  { {d{\omega _1}} } {\left| {f({\omega _1})} \right|^2}[1 + \cos({\omega _1}{T_1})].\\
\end{aligned}
\end{array}
\end{equation}
In the above calculation, the relationship of $\frac{1}{2 \pi} \int_{-\infty}^{\infty} e^{-i(\omega-\omega^{\prime}) t} d t=\delta(\omega-\omega^{\prime})$ is utilized.

\subsection*{Appendix 2: Calculation of folded Franson interference}
Then, we deduce the equations for the folded Franson interference using multi-mode theory. The setup of the Franson interference is shown in Fig.\,\ref{SupplementalFig}(b).
The two-photon state from a spontaneous parametric down-conversion (SPDC) process can be described as
\begin{equation}\label{eqA1}
|\psi\rangle=\int_0^{\infty} d \omega_s \int_0^{\infty} d \omega_i f\left(\omega_s, \omega_i\right) \hat{a}_{s H}^{\dagger}\left(\omega_s\right) \hat{a}_{i V}^{\dagger}\left(\omega_i\right)|00\rangle,
\end{equation}
where $\omega$ is the angular frequency; $\hat a^\dag$ is the creation operator and the subscripts $s$ and $i$ denote the signal and idler photons from SPDC, respectively; $H$ and $V$ represent the polarization of signal and idler photons; $f(\omega _s ,\omega _i )$ is the joint spectral amplitude of the signal and idler photons.
The detection field operators of detector 1 (D5) and detector 2 (D6) are
\begin{equation}\label{eqA2}
\hat E_5^{( + )} (t_5 ) = \frac{1}{{\sqrt {2\pi } }}\int_0^\infty  {d\omega _5 } \hat a_5 (\omega _5 )e^{ - i\omega _5 t_5 },
\end{equation}
\begin{equation}\label{eqA3}
\hat E_6^{( + )} (t_6 ) = \frac{1}{{\sqrt {2\pi } }}\int_0^\infty  {d\omega _6 \hat a_6 (\omega _6 )} e^{ - i\omega _6 t_6 },
\end{equation}
where the subscripts $5$ and $6$ denote the photons detected by D5 and D6 respectively.
The transformation rule after the delay time $\tau$ is
\begin{equation}\label{eqA4}
\begin{aligned}
& \hat{a}_5\left(\omega_5\right)=\frac{1}{\sqrt{2}} \hat{a}_4\left(\omega_5\right)=\frac{1}{2}\left[\hat{a}_3\left(\omega_5\right) e^{-i \omega_5 \tau}+\hat{a}_2\left(\omega_5\right)\right]=\frac{1}{2 \sqrt{2}}\left[\hat{a}_1\left(\omega_5\right) e^{-i \omega_5 \tau}+\hat{a}_1\left(\omega_5\right)\right]=\frac{1}{2 \sqrt{2}}\left(e^{-i \omega_5 \tau}+1\right) \hat{a}_1\left(\omega_5\right) \\
\end{aligned},
\end{equation}
\begin{equation}\label{eqA5}
\begin{aligned}
& \hat{a}_6\left(\omega_6\right)=\frac{1}{\sqrt{2}} \hat{a}_4\left(\omega_6\right)=\frac{1}{2}\left[\hat{a}_3\left(\omega_6\right) e^{-i \omega_6 \tau}+\hat{a}_2\left(\omega_6\right)\right]=\frac{1}{2 \sqrt{2}}\left[\hat{a}_1\left(\omega_6\right) e^{-i \omega_6 \tau}+\hat{a}_1\left(\omega_6\right)\right]=\frac{1}{2 \sqrt{2}}\left(e^{-i \omega_6 \tau}+1\right) \hat{a}_1\left(\omega_6\right) \\
\end{aligned}.
\end{equation}

So, we can rewrite the field operators as
\begin{equation}\label{eqA6}
\begin{aligned}
& \hat{E}_5^{(+)}\left(t_5\right)=\frac{1}{\sqrt{2 \pi}} \int_0^{\infty} d \omega_5 \hat{a}_5\left(\omega_5\right) e^{-i \omega_5 t_5}=\frac{1}{4 \sqrt{\pi}} \int_0^{\infty} d \omega_5\left(e^{-i \omega_5 \tau}+1\right) \hat{a}_1\left(\omega_5\right) e^{-i \omega_5 t_5} \\
\end{aligned},\\
\end{equation}
and
\begin{equation}\label{eq7}
\begin{aligned}
& \hat{E}_6^{(+)}\left(t_6\right)=\frac{1}{\sqrt{2 \pi}} \int_0^{\infty} d \omega_6 \hat{a}_6\left(\omega_6\right) e^{-i \omega_6 t_6}=\frac{1}{4 \sqrt{\pi}} \int_0^{\infty} d \omega_6\left(e^{-i \omega_6 \tau}+1\right) \hat{a}_1\left(\omega_6\right) e^{-i \omega_6 t_6} \\
\end{aligned}.\\
 \end{equation}
 
Consider the polarization:
\begin{equation}
\hat{E}_5^{(+)}\left(t_5\right) \hat{E}_6^{(+)}\left(t_6\right)=\hat{E}_{5 H}^{(+)}\left(t_5\right) \hat{E}_{6 V}^{(+)}\left(t_6\right)+\hat{E}_{5 V}^{(+)}\left(t_5\right) \hat{E}_{6 H}^{(+)}\left(t_6\right)+\hat{E}_{5 H}^{(+)}\left(t_5\right) \hat{E}_{6 H}^{(+)}\left(t_6\right)+\hat{E}_{5 V}^{(+)}\left(t_5\right) \hat{E}_{6 V}^{(+)}\left(t_6\right).
\end{equation}

In the above equation, only 2 out of 4 terms exist:
\begin{equation}\label{eq7}
\hat{E}_5^{(+)}\left(t_5\right) \hat{E}_6^{(+)}\left(t_6\right)=\hat{E}_{5 H}^{(+)}\left(t_5\right) \hat{E}_{6 V}^{(+)}\left(t_6\right)+\hat{E}_{5 V}^{(+)}\left(t_5\right) \hat{E}_{6 H}^{(+)}\left(t_6\right).
\end{equation}

The coincidence probability $P(\tau )$, which is also the second-order correlation function $G_2(\tau)$, can be expressed as
\begin{equation}
\begin{aligned}
P(\tau ) \equiv G_2(\tau) & =\int_{-\infty}^{\infty} \int_{-\infty}^{\infty} d t_5 d t_6\left\langle\psi\left|\hat{E}_6^{(-)} \hat{E}_5^{(-)} \hat{E}_5^{(+)} \hat{E}_6^{(+)}\right| \psi\right\rangle \\ & =\int_{-\infty}^{\infty} \int_{-\infty}^{\infty} d t_5 d t_6\left\langle\psi\left|\hat{E}_{6 V}^{(-)} \hat{E}_{5 H}^{(-)} \hat{E}_{5 H}^{(+)} \hat{E}_{6 V}^{(+)}\right| \psi\right\rangle+\int_{-\infty}^{\infty} \int_{-\infty}^{\infty} d t_5 d t_6\left\langle\psi\left|\hat{E}_{6 H}^{(-)} \hat{E}_{5 V}^{(-)} \hat{E}_{5 V}^{(+)} \hat{E}_{6 H}^{(+)}\right| \psi\right\rangle \\ & =P_{H V}(\tau)+P_{V H}(\tau).\end{aligned}\\
\end{equation}

First of all, consider $P_{H V}(\tau)=\int_{-\infty}^{\infty} \int_{-\infty}^{\infty} d t_5 d t_6\left\langle\psi\left|\hat{E}_{6 V}^{(-)} \hat{E}_{5 H}^{(-)} \hat{E}_{5 H}^{(+)} \hat{E}_{6 V}^{(+)}\right| \psi\right\rangle$. In this equation:
\begin{equation}
\begin{aligned} \hat{E}_{5 H}^{(+)}\left(t_5\right) \hat{E}_{6 V}^{(+)}\left(t_6\right) & =\frac{1}{4 \sqrt{\pi}} \int_0^{\infty} d \omega_5\left(e^{-i \omega_5 \tau}+1\right) \hat{a}_{1 H}\left(\omega_5\right) e^{-i \omega_5 t_5} \times \frac{1}{4 \sqrt{\pi}} \int_0^{\infty} d \omega_6\left(e^{-i \omega_6 \tau}+1\right) \hat{a}_{1 V}\left(\omega_6\right) e^{-i \omega_6 t_6} \\ & =\frac{1}{16 \pi} \int_0^{\infty} \int_0^{\infty} d \omega_5 d \omega_6\left(e^{-i \omega_5 \tau}+1\right)\left(e^{-i \omega_6 \tau}+1\right) \hat{a}_{1 H}\left(\omega_5\right) \hat{a}_{1 V}\left(\omega_6\right) e^{-i \omega_5 t_5} e^{-i \omega_6 t_6}.\end{aligned}
\end{equation}

Then, consider $\hat{E}_{5 H}^{(+)}\left(t_5\right) \hat{E}_{6 V}^{(+)}\left(t_6\right)|\psi\rangle$
\begin{equation}
\begin{aligned}
\hat{E}_{5 H}^{(+)}\left(t_5\right) \hat{E}_{6 V}^{(+)}\left(t_6\right)|\psi\rangle & =\frac{1}{16 \pi} \int_0^{\infty} \int_0^{\infty} d \omega_5 d \omega_6\left(e^{-i \omega_5 \tau}+1\right)\left(e^{-i \omega_6 \tau}+1\right) \hat{a}_{1 H}\left(\omega_5\right) \hat{a}_{1 V}\left(\omega_6\right) e^{-i \omega_5 t_5} e^{-i \omega_6 t_6} \\
& \times \int_0^{\infty} d \omega_s \int_0^{\infty} d \omega_i f\left(\omega_s, \omega_i\right) \hat{a}_{s H}^{\dagger}\left(\omega_s\right) \hat{a}_{i V}^{\dagger}\left(\omega_i\right)|00\rangle \\
& =\frac{1}{16 \pi} \int_0^{\infty} \int_0^{\infty} d \omega_5 d \omega_6 \int_0^{\infty} \int_0^{\infty} d \omega_s d \omega_i \delta\left(\omega_5-\omega_s\right) \delta\left(\omega_6-\omega_i\right) \\
& \times f\left(\omega_s, \omega_i\right)\left(e^{-i \omega_5 \tau}+1\right)\left(e^{-i \omega_6 \tau}+1\right) e^{-i \omega_5 t_5} e^{-i \omega_6 t_6}|00\rangle \\
& =\frac{1}{16 \pi} \int_0^{\infty} \int_0^{\infty} d \omega_5 d \omega_6 f\left(\omega_5, \omega_6\right)\left(e^{-i \omega_5 \tau}+1\right)\left(e^{-i \omega_6 \tau}+1\right) e^{-i \omega_5 t_5} e^{-i \omega_6 t_6}|00\rangle.
\end{aligned}
\end{equation}

In the above calculation, the equations of $ \hat{a}_{1 H}\left(\omega_5\right) \hat{a}_{s H}^{\dagger}\left(\omega_s\right)|0\rangle=\delta\left(\omega_5-\omega_s\right)|0\rangle $ and $\hat{a}_{1 V}\left(\omega_6\right) \hat{a}_{i V}^{\dagger}\left(\omega_i\right)|0\rangle=\delta\left(\omega_6-\omega_i\right)|0\rangle
$ are used. $\hat{a}_{s H}^{\dagger}\left(\omega_s\right)$ and $\hat{a}_{1 H}$ are both acting on H photons in path 1, so $\hat{a}_{s H}^{\dagger} \equiv \hat{a}_{1 H}$.

Then,
\begin{equation}
\begin{aligned}
\left\langle\psi\left|\hat{E}_{6 V}^{(-)} \hat{E}_{5 H}^{(-)} \hat{E}_{5 H}^{(+)} \hat{E}_{6 V}^{(+)}\right| \psi\right\rangle & =\frac{1}{16 \pi} \int_0^{\infty} \int_0^{\infty} d \omega_5^{\prime} d \omega_6^{\prime} f^*\left(\omega_5^{\prime}, \omega_6^{\prime}\right)\left(e^{i \omega_5^{\prime} \tau}+1\right)\left(e^{i \omega_6^{\prime} \tau}+1\right) e^{i \omega_5^{\prime} t_5} e^{i \omega_6^{\prime} t_6} \\
& \times \frac{1}{16 \pi} \int_0^{\infty} \int_0^{\infty} d \omega_5 d \omega_6 f\left(\omega_5, \omega_6\right)\left(e^{-i \omega_5 \tau}+1\right)\left(e^{-i \omega_6 \tau}+1\right) e^{-i \omega_5 t_5} e^{-i \omega_6 t_6} \\
& =\frac{1}{256\pi^2}  \int_0^{\infty} \int_0^{\infty} d \omega_5 d \omega_6 \int_0^{\infty} \int_0^{\infty} d \omega_5^{\prime} d \omega_6^{\prime} f\left(\omega_5, \omega_6\right) f^*\left(\omega_5^{\prime}, \omega_6^{\prime}\right) \\
& \times\left(e^{-i \omega_5 \tau}+1\right)\left(e^{-i \omega_6 \tau}+1\right) e^{-i \omega_5 t_5} e^{-i \omega_6 t_6}\left(e^{i \omega_5^{\prime} \tau}+1\right)\left(e^{i \omega_6^{\prime} \tau}+1\right) e^{i \omega_5^{\prime} t_5} e^{i \omega_6^{\prime} t_6}.
\end{aligned}
\end{equation}

Finally,
\begin{equation}
\begin{aligned}
P_{H V}(\tau) & =\int_{-\infty}^{\infty} \int_{-\infty}^{\infty} d t_5 d t_6\left\langle\psi\left|\hat{E}_{6 V}^{(-)} \hat{E}_{5 H}^{(-)} \hat{E}_{5 H}^{(+)} \hat{E}_{6 V}^{(+)}\right| \psi\right\rangle \\
& =\frac{1}{256 \pi^2} \int_{-\infty}^{\infty} \int_{-\infty}^{\infty} d t_5 d t_6 \int_0^{\infty} \int_0^{\infty} d \omega_5 d \omega_6 \int_0^{\infty} \int_0^{\infty} d \omega_5^{\prime} d \omega_6^{\prime} f\left(\omega_5, \omega_6\right) f^*\left(\omega_5^{\prime}, \omega_6^{\prime}\right) \\
& \times\left(e^{-i \omega_5 \tau}+1\right)\left(e^{-i \omega_6 \tau}+1\right) e^{-i \omega_5 t_5} e^{-i \omega_6 t_6}\left(e^{i \omega_5^{\prime} \tau}+1\right)\left(e^{i \omega_6^{\prime} \tau}+1\right) e^{i \omega_5^{\prime} t_5} e^{i \omega_6^{\prime} t_6}.
\end{aligned}
\end{equation}

By utilizing $\frac{1}{2 \pi} \int_{-\infty}^{\infty} e^{-i(\omega-\omega^{\prime}) t} d t=\delta(\omega-\omega^{\prime})$, the above equation can be further simplified as \begin{equation}\label{eq0}
\begin{aligned}
P_{H V}(\tau) & =\int_{-\infty}^{\infty} \int_{-\infty}^{\infty} d t_5 d t_6\left\langle\psi\left|\hat{E}_{6 V}^{(-)} \hat{E}_{5 H}^{(-)} \hat{E}_{5 H}^{(+)} \hat{E}_{6 V}^{(+)}\right| \psi\right\rangle \\
& =\frac{1}{64} \int_0^{\infty} \int_0^{\infty} \int_0^{\infty} \int_0^{\infty} d \omega_5 d \omega_6 d \omega_5^{\prime} d \omega_6^{\prime} \delta\left(\omega_5-\omega_5^{\prime}\right) \delta\left(\omega_6-\omega_6^{\prime}\right) f\left(\omega_5, \omega_6\right) \\
& \times\left(e^{-i \omega_5 \tau}+1\right)\left(e^{-i \omega_6 \tau}+1\right) f^*\left(\omega_5^{\prime}, \omega_6^{\prime}\right)\left(e^{i \omega_5^{\prime} \tau}+1\right)\left(e^{i \omega_6^{\prime} \tau}+1\right) \\
& =\frac{1}{64} \int_0^{\infty} \int_0^{\infty} d \omega_5 d \omega_6 f\left(\omega_5, \omega_6\right) f^*\left(\omega_5, \omega_6\right)\left(e^{-i \omega_5 \tau}+1\right)\left(e^{-i \omega_6 \tau}+1\right)\left(e^{i \omega_5^{\prime} \tau}+1\right)\left(e^{i \omega_6^{\prime} \tau}+1\right) \\
& =\frac{1}{64} \int_0^{\infty} \int_0^{\infty} d \omega_5 d \omega_6\left|f\left(\omega_5, \omega_6\right)\right|^2\left|\left(e^{-i \omega_5 \tau}+1\right)\left(e^{-i \omega_6 \tau}+1\right)\right|^2 \\
& =\frac{1}{16} \int_0^{\infty} \int_0^{\infty} d \omega_5 d \omega_6\left|f\left(\omega_5, \omega_6\right)\right|^2\left[1+\operatorname{cos}\left(\omega_5 \tau\right)\right]\left[1+\operatorname{cos}\left(\omega_6 \tau\right)\right].
\end{aligned}
\end{equation}

Similarly,
\begin{equation}
\begin{aligned}
P_{V H}(\tau) & =\int_{-\infty}^{\infty} \int_{-\infty}^{\infty} d t_5 d t_6\left\langle\psi\left|\hat{E}_{6 H}^{(-)} \hat{E}_{5 V}^{(-)} \hat{E}_{5 V}^{(+)} \hat{E}_{6 H}^{(+)}\right| \psi\right\rangle \\
& =\frac{1}{64} \int_0^{\infty} \int_0^{\infty} \int_0^{\infty} \int_0^{\infty} d \omega_5 d \omega_6 d \omega_5^{\prime} d \omega_6^{\prime} \delta\left(\omega_5-\omega_5^{\prime}\right) \delta\left(\omega_6-\omega_6^{\prime}\right) f\left(\omega_6, \omega_5\right) \\
& \times\left(e^{-i \omega_5 \tau}+1\right)\left(e^{-i \omega_6 \tau}+1\right) f^*\left(\omega_6^{\prime}, \omega_5^{\prime}\right)\left(e^{i \omega_5^{\prime} \tau}+1\right)\left(e^{i \omega_6^{\prime} \tau}+1\right) \\
& =\frac{1}{64} \int_0^{\infty} \int_0^{\infty} d \omega_5 d \omega_6 f\left(\omega_6, \omega_5\right) f^*\left(\omega_6, \omega_5\right)\left(e^{-i \omega_5 \tau}+1\right)\left(e^{-i \omega_6 \tau}+1\right)\left(e^{i \omega_5^{\prime} \tau}+1\right)\left(e^{i \omega_6^{\prime} \tau}+1\right) \\
& =\frac{1}{64} \int_0^{\infty} \int_0^{\infty} d \omega_5 d \omega_6\left|f\left(\omega_6, \omega_5\right)\right|^2\left|\left(e^{-i \omega_5 \tau}+1\right)\left(e^{-i \omega_6 \tau}+1\right)\right|^2 \\
& =\frac{1}{16} \int_0^{\infty} \int_0^{\infty} d \omega_5 d \omega_6\left|f\left(\omega_6, \omega_5\right)\right|^2\left[1+\operatorname{cos}\left(\omega_5 \tau\right)\right]\left[1+\operatorname{cos}\left(\omega_6 \tau\right)\right].
\end{aligned}
\end{equation}

Finally, the coincidence probability $P(\tau )$ is:
\begin{equation}
\begin{aligned}
P(\tau) & =P_{H V}(\tau)+P_{V H}(\tau) \\
& =\frac{1}{16} \int_0^{\infty} \int_0^{\infty} d \omega_5 d \omega_6\left|f\left(\omega_5, \omega_6\right)\right|^2\left[1+\operatorname{cos}\left(\omega_5 \tau\right)\right]\left[1+\operatorname{cos}\left(\omega_6 \tau\right)\right] \\
& +\frac{1}{16} \int_0^{\infty} \int_0^{\infty} d \omega_5 d \omega_6\left|f\left(\omega_6, \omega_5\right)\right|^2\left[1+\operatorname{cos}\left(\omega_5 \tau\right)\right]\left[1+\operatorname{cos}\left(\omega_6 \tau\right)\right] \\
& =\frac{1}{16} \int_0^{\infty} \int_0^{\infty} d \omega_5 d \omega_6\left[\left|f\left(\omega_5, \omega_6\right)\right|^2+\left|f\left(\omega_6, \omega_5\right)\right|^2\right]\left[1+\operatorname{cos}\left(\omega_5 \tau\right)\right]\left[1+\operatorname{cos}\left(\omega_6 \tau\right)\right].
\end{aligned}
\end{equation}

If $f\left(\omega_5, \omega_6\right)=f\left(\omega_5, \omega_6\right)$, then the coincidence probability $P(\tau )$ is
\begin{equation}
\begin{aligned}
P(\tau)  =\frac{1}{8} \int_0^{\infty} \int_0^{\infty} d \omega_5 d \omega_6 \left|f\left(\omega_5, \omega_6\right)\right|^2\left[1+\operatorname{cos}\left(\omega_5 \tau\right)\right]\left[1+\operatorname{cos}\left(\omega_6 \tau\right)\right].
\end{aligned}
\end{equation}

\subsection*{Appendix 3: Calculation single counts of folded Franson interference}
In this section, we deduce the single counts equations for the Franson interference using multi-mode theory. The setup of the Franson interference is shown in Fig.\,\ref{SupplementalFig}.
The two-photon state from a spontaneous parametric down-conversion (SPDC) process can be described as
\begin{equation}\label{eqA1}
\left| \psi  \right\rangle  = \int_0^{\infty}{\int_0^\infty  {d\omega _s d\omega _i } } f(\omega _s ,\omega _i )\hat a_s^\dag  (\omega _s )\hat a_i^\dag  (\omega _i )\left| {00} \right\rangle,
\end{equation}
where $\omega$ is the angular frequency; $\hat a^\dag$ is the creation operator and the subscripts $s$ and $i$ denote the signal and idler photons from SPDC,  respectively; $f(\omega _s ,\omega _i )$ is the joint spectral amplitude of the signal and idler photons.
where $\omega$ is the angular frequency; $\hat a^\dag$ is the creation operator and the subscripts $s$ and $i$ denote the signal and idler photons from SPDC, respectively; $H$ and $V$ represent the polarization of signal and idler photons; and $f(\omega _s ,\omega _i )$ is the joint spectral amplitude of the signal and idler photons.
The detection field operators of detector 1 (D5) and detector 2 (D6) are
\begin{equation}\label{eqA2}
\hat E_5^{( + )} (t_5 ) = \frac{1}{{\sqrt {2\pi } }}\int_0^\infty  {d\omega _5 } \hat a_5 (\omega _5 )e^{ - i\omega _5 t_5 },
\end{equation}
\begin{equation}\label{eqA3}
\hat E_6^{( + )} (t_6 ) = \frac{1}{{\sqrt {2\pi } }}\int_0^\infty  {d\omega _6 \hat a_6 (\omega _6 )} e^{ - i\omega _6 t_6 },
\end{equation}
where the subscripts $5$ and $6$ denote the photons detected by D5 and D6 respectively.
The transformation rule after the delay time $T$ is (take D5 for example)
\begin{equation}\label{eqA4}
\begin{aligned}
& \hat{a}_5\left(\omega_5\right)=\frac{1}{\sqrt{2}} \hat{a}_4\left(\omega_5\right)=\frac{1}{2}\left[\hat{a}_3\left(\omega_5\right) e^{-i \omega_5 \tau}+\hat{a}_2\left(\omega_5\right)\right]=\frac{1}{2 \sqrt{2}}\left[\hat{a}_1\left(\omega_5\right) e^{-i \omega_5 \tau}+\hat{a}_1\left(\omega_5\right)\right]=\frac{1}{2 \sqrt{2}}\left(e^{-i \omega_5 \tau}+1\right) \hat{a}_1\left(\omega_5\right). \\
\end{aligned}
\end{equation}

So, we can rewrite the field operators as
\begin{equation}\label{eqA6}
\begin{aligned}
& \hat{E}_5^{(+)}\left(t_5\right)=\frac{1}{\sqrt{2 \pi}} \int_0^{\infty} d \omega_5 \hat{a}_5\left(\omega_5\right) e^{-i \omega_5 t_5}=\frac{1}{4 \sqrt{\pi}} \int_0^{\infty} d \omega_5\left(e^{-i \omega_5 \tau}+1\right) \hat{a}_1\left(\omega_5\right) e^{-i \omega_5 t_5}. \\
\end{aligned}\\
\end{equation}

The single counts' probability $P(\tau )$, can be expressed as
\begin{equation}\label{eqA1}
P(\tau)=P_H(\tau)+P_V(\tau)=\int_{-\infty}^{\infty} d t_5\left\langle\psi\left|\hat{E}_{5 H}^{(-)}\left(t_5\right) \hat{E}_{5 H}^{(+)}\left(t_5\right)\right| \psi\right\rangle+\int_{-\infty}^{\infty} d t_5\left\langle\psi\left|\hat{E}_{5 V}^{(-)}\left(t_5\right) \hat{E}_{5 V}^{(+)}\left(t_5\right)\right| \psi\right\rangle.
\end{equation}

First of all, consider$\hat{E}_{5 H}^{(+)}\left(t_5\right)|\psi\rangle$
\begin{equation}\label{eqA6}
\begin{aligned}
& \hat{E}_{5 H}^{(+)}\left(t_5\right)|\psi\rangle=\frac{1}{4 \sqrt{\pi}} \int_0^{\infty} d \omega_5\left(e^{-i \omega_5 \tau}+1\right) \hat{a}_{1 H}\left(\omega_5\right) e^{-i \omega_5 t_5} \times \int_0^{\infty} d \omega_s \int_0^{\infty} d \omega_i f\left(\omega_s, \omega_i\right) \hat{a}_{s H}^{\dagger}\left(\omega_s\right) \hat{a}_{i V}^{\dagger}\left(\omega_i\right)|00\rangle \\
& =\frac{1}{4 \sqrt{\pi}} \int_0^{\infty} d \omega_5 \int_0^{\infty} d \omega_s \int_0^{\infty} d \omega_i f\left(\omega_s, \omega_i\right)\left(e^{-i \omega_5 \tau}+1\right) e^{-i \omega_5 t_5} \delta\left(\omega_s-\omega_5\right) \hat{a}_{i V}^{\dagger}\left(\omega_i\right)|00\rangle \\
& =\frac{1}{4 \sqrt{\pi}} \int_0^{\infty} d \omega_s \int_0^{\infty} d \omega_i f\left(\omega_s, \omega_i\right)\left(e^{-i \omega_s \tau}+1\right) e^{-i \omega_s t_5} \hat{a}_{i V}^{\dagger}\left(\omega_i\right)|00\rangle ,\\
& \left\langle\psi\left|\hat{E}_{5 H}^{(-)}\left(t_5\right) \hat{E}_{5 H}^{(+)}\left(t_5\right)\right| \psi\right\rangle \\
& =\frac{1}{16 \pi} \int_0^{\infty} d \omega_s^{\prime} \int_0^{\infty} d \omega_i^{\prime} f^*\left(\omega_s^{\prime}, \omega_i^{\prime}\right)\left(e^{i \omega_s^{\prime} \tau}+1\right) e^{i \omega_s^{\prime} t_5} \hat{a}_{i V}\left(\omega_i^{\prime}\right) \times \int_0^{\infty} d \omega_s \int_0^{\infty} d \omega_i f\left(\omega_s, \omega_i\right)\left(e^{-i \omega_s \tau}+1\right) e^{-i \omega_s t_5} \hat{a}_{i V}^{\dagger}\left(\omega_i\right) \\
& =\frac{1}{16 \pi} \int_0^{\infty} d \omega_s^{\prime} \int_0^{\infty} d \omega_i^{\prime} f^*\left(\omega_s^{\prime}, \omega_i^{\prime}\right)\left(e^{i \omega_s^{\prime} \tau}+1\right) e^{i \omega_s^{\prime} t_5} \times \int_0^{\infty} d \omega_s \int_0^{\infty} d \omega_i f\left(\omega_s, \omega_i\right)\left(e^{-i \omega_s \tau}+1\right) e^{-i \omega_s t_5} \delta\left(\omega_i-\omega_i^{\prime}\right) \\
& =\frac{1}{16 \pi} \int_0^{\infty} d \omega_s^{\prime} f^*\left(\omega_s^{\prime}, \omega_i\right)\left(e^{i \omega_s^{\prime} \tau}+1\right) e^{i \omega_s^{\prime} t_5} \times \int_0^{\infty} d \omega_s \int_0^{\infty} d \omega_i f\left(\omega_s, \omega_i\right)\left(e^{-i \omega_s \tau}+1\right) e^{-i \omega_s t_5} \\
& =\frac{1}{16\pi}  \int_0^{\infty} d \omega_s \int_0^{\infty} d \omega_i \int_0^{\infty} d \omega_s^{\prime} f^*\left(\omega_s^{\prime}, \omega_i\right) f\left(\omega_s, \omega_i\right)\left(e^{i \omega_s^{\prime} \tau}+1\right)\left(e^{-i \omega_s \tau}+1\right) e^{i \omega_s^{\prime} t_5} e^{-i \omega_s t_5}. \\
&
\end{aligned}
\end{equation}

Then,
\begin{equation}
\begin{aligned}
& P_H(\tau)=\int_{-\infty}^{\infty} d t_5\left\langle\psi\left|\hat{E}_{5 H}^{(-)}\left(t_5\right) \hat{E}_{5 H}^{(+)}\left(t_5\right)\right| \psi\right\rangle \\
& =\frac{1}{8} \int_0^{\infty} d \omega_s \int_0^{\infty} d \omega_i \int_0^{\infty} d \omega_s^{\prime} f^*\left(\omega_s^{\prime}, \omega_i\right) f\left(\omega_s, \omega_i\right)\left(e^{i \omega_s^{\prime} \tau}+1\right)\left(e^{-i \omega_s \tau}+1\right) \delta\left(\omega_s-\omega_s^{\prime}\right) \\
& =\frac{1}{8} \int_0^{\infty} d \omega_s \int_0^{\infty} d \omega_i f^*\left(\omega_s, \omega_i\right) f\left(\omega_s, \omega_i\right)\left(e^{i \omega_s \tau}+1\right)\left(e^{-i \omega_s \tau}+1\right) \\
& =\frac{1}{8} \int_0^{\infty} d \omega_s \int_0^{\infty} d \omega_i\left|f\left(\omega_s, \omega_i\right)\left(e^{-i \omega_s \tau}+1\right)\right|^2.
\end{aligned}
\end{equation}

Similarly,
\begin{equation}
\begin{aligned}
& P_V(\tau)=\int_{-\infty}^{\infty} d t_5\left\langle\psi\left|\hat{E}_{5 V}^{(-)}\left(t_5\right) \hat{E}_{5 V}^{(+)}\left(t_5\right)\right| \psi\right\rangle \\
& =\frac{1}{8} \int_0^{\infty} d \omega_i^{\prime} \int_0^{\infty} d \omega_s \int_0^{\infty} d \omega_i f^*\left(\omega_s, \omega_i^{\prime}\right) f\left(\omega_s, \omega_i\right)\left(e^{i \omega_i^{\prime} \tau}+1\right)\left(e^{-i \omega_i \tau}+1\right) \delta\left(\omega_i^{\prime}-\omega_i\right) \\
& =\frac{1}{8} \int_0^{\infty} d \omega_i^{\prime} \int_0^{\infty} d \omega_s f^*\left(\omega_s, \omega_i \right) f\left(\omega_s, \omega_i\right)\left(e^{i \omega_i \tau}+1\right)\left(e^{-i \omega_i \tau}+1\right) \\
& =\frac{1}{8} \int_0^{\infty} d \omega_s \int_0^{\infty} d \omega_i\left|f\left(\omega_s, \omega_i\right)\left(e^{-i \omega_i \tau}+1\right)\right|^2.
\end{aligned}
\end{equation}

Finally, the single counts' probability $P(\tau )$ is:
\begin{equation}
\begin{aligned}
& P(\tau)=P_H(\tau)+P_V(\tau)=\frac{1}{8} \int_{0}^{\infty} d \omega_s \int_{0}^{\infty} d \omega_i\left|f\left(\omega_s, \omega_i\right)\left(e^{-i \omega_s \tau}+1\right)\right|^2+\frac{1}{8} \int_{0}^{\infty} d \omega_s \int_{0}^{\infty} d \omega_i\left|f\left(\omega_s, \omega_i\right)\left(e^{-i \omega_i \tau}+1\right)\right|^2 \\
& =\frac{1}{4} \int_0^{\infty} d \omega_s \int_0^{\infty} d \omega_i\left|f\left(\omega_s, \omega_i\right)\right|^2\left[1+\operatorname{cos}\left(\omega_s \tau\right)+1+\operatorname{cos}\left(\omega_i \tau\right)\right].
\end{aligned}
\end{equation}

\end{document}